\documentclass[12pt]{article}
\usepackage{graphicx,latexsym,natbib}
\newcommand{\ndef}[1]{\emph{#1}}

\newcommand{\myfig}[2]{\begin{center}
  \includegraphics[scale=#1]{#2.ps}
   \end{center} }
\newcommand{\set}[1]{\left\{#1\right\}}
\newcommand{\To}{\rightarrow}

\newcommand{\ZZ}{{\mathbf{Z}}}
\newcommand{\NN}{{\mathbf{N}}}

\title{Modeling diffusion of innovations
with probabilistic cellular automata}
\author{Nino Boccara and Henryk Fuk\'s}
\begin{document}
\maketitle
\noindent
{\bf Abstract:} We present a family  of one-dimensional cellular automata
modeling the diffusion of an innovation in a population. Starting from
simple deterministic rules, we
construct models parameterized by the interaction range and
exhibiting a second-order phase transition. We show that the number
of individuals who eventually keep adopting the
innovation strongly depends on connectivity between individuals.

\section{Introduction}
Diffusion phenomena in social systems such as spread of news, rumors
or innovations have been extensively studied for the past three
decades by social scientists, geographers, economists, as well as
management and marketing scholars. Traditionally, ordinary
differential equations have been used to model these phenomena,
beginning with the Bass model \citep{bass69} and ending with
sophisticated models that take into account learning, risk aversion,
nature of innovation, etc. \citep{Mahajan1990, Mahajan85,Rogers95}.
Models incorporating space and spatial distribution of individuals
have been also proposed, although most research in this field has been
directed to refining of the discrete H{\"a}gerstrand models
\citep{Hagerstrand52,Hagerstrand65} and to constructing partial
differential equations similar to diffusion equations known to
physicists \citep{Haynes77}.

Diffusion of innovations (we will use this term in a general sense,
meaning also news, rumors, new products,
etc.) is usually defined as a process by which the innovation ``is
communicated through certain channels over time among the members of a
social system'' \citep{Rogers95}. In most cases, these communication
channels have a rather short range, i.e. in our decisions we are heavily
influenced by our friends, family, coworkers, but not that much by
unknown people in distant cities. This local nature of social
interactions makes cellular automata (CA) a well-adapted tool in
modeling diffusion phenomena. In fact, epidemic models formulated
in terms of automata networks have been successfully constructed in
recent years \citep{bc92,bc93,bco94}.

\section{Simple deterministic models}
\label{sec:deterinno}
We will  construct  models of diffusion of innovations
based on elementary (radius-1) cellular automata \citep{Wolfram86}.
If $s(i,t)$ denotes the state of lattice site $i$ at time $t$, the
function $(i,t) \mapsto s(i,t)$ is a mapping from $\ZZ \times \NN$ to 
$\ZZ_2=\{0,1\}$. Given a
function $f:\{0,1\}^{2r+1}\mapsto\{0,1\}$, the discrete dynamical system
\begin{equation}
s(i,t+1)=f(s(i-r,t),s(i-r+1,t), \ldots, s(i+r,t))
\end{equation}
is called a cellular automaton (CA) of radius $r$, and $f$ is 
called its local (function) rule.

In the simplest version of our model, the sites of an infinite lattice
are all occupied by individuals. The individuals are of two different
types: {\em adopters} $\Big( s(i,t)=1 \Big)$ and {\em neutrals}
$\Big( s(i,t)=0 \Big)$.
Moreover, we wil assume that an individual can get information only from
it
two nearest neighbors. To simplify our model, we will also assume that
once an individual becomes an adopter, he remains an adopter, i.e. his
state cannot change. This condition fixes four entries in the rule
table: $\set{0,1,0} \To 1$, $\set{0,1,1} \To 1$, $\set{1,1,0} \To 1$
and $\set{1,1,1} \To 1$, and since the information comes from nearest
neighbors, if both are neutral, then the individual will stay neutral,
i.e., $\set{0,0,0} \To 0$. This leaves three entries in the rule
table to be determined,
and this can be done in $2^3=8$ ways, as shown in
Table~\ref{tab:eightrules}. The first
\begin{table}
\begin{center}
\begin{tabular}{|c||c|c|c|c|c|c|c|c|} \hline
 code&1,1,1&1,1,0&1,0,1&1,0,0&0,1,1&0,1,0&0,0,1&0,0,0 \\ \hline \hline
  204&  1  &  1  &  0  &  0  &  1  &  1  &  0  &  0   \\
  206&  1  &  1  &  0  &  0  &  1  &  1  &  1  &  0   \\
  220&  1  &  1  &  0  &  1  &  1  &  1  &  0  &  0   \\
  222&  1  &  1  &  0  &  1  &  1  &  1  &  1  &  0   \\
  236&  1  &  1  &  1  &  0  &  1  &  1  &  0  &  0   \\
  238&  1  &  1  &  1  &  0  &  1  &  1  &  1  &  0   \\
  252&  1  &  1  &  1  &  1  &  1  &  1  &  0  &  0   \\
  254&  1  &  1  &  1  &  1  &  1  &  1  &  1  &  0   \\ \hline
\end{tabular}
\end{center}
\caption{Eight possible elementary rules for diffusion of an innovation.}
\label{tab:eightrules}
\end{table}
Rule $204$ (for rule codes cf. Wolfram 86)  listed in this table is just
the identity, and it will be
excluded  from further considerations. Rules $220$ and $252$ can be
obtained respectively from $206$ and $238$ by spatial reflection,
therefore they will be excluded  too. This leaves us with five
distinct rules.
\begin{description}
 \item[Rule 254:] An individual adopts if, at least, one of his neighbors
is an adopter.
 \item[Rule 238:] An individual adopts only if  his right neighbor
 is an adopter.
 \item[Rule 222:] An individual adopts only if exactly one of his neighbors
 is an adopter.
 \item[Rule 206:] An individual adopts only if his right neighbor
 is an adopter and its left neighbor is neutral.
 \item[Rule 236:] An individual adopts only if both his neighbors are
adopters.
\end{description}
We will now demonstrate that in all five cases, the density of
adopters $\rho(t)$ at time $t$ can be exactly computed, assuming that
we start from a disordered initial configuration with $\rho(0)=\rho_0$.

(i) In order to understand the dynamics of Rule 254, we can view the
initial configuration as clusters of ones separated by clusters of
zeros. Only neutral sites adjacent to a cluster of ones change their
state, while all other neutral sites remain neutral, as shown in
the example below (sites that will change their state to  $1$ in the
next time step are underlined):
\begin{center}
$\cdots$
\begin{tabular}{|c|c|c|c|c|c|c|c|c|c|c|c|}\hline
                 1&\underline{0}&0&\underline{0}&1&1&\underline{0}&0&0&\underline{0}&1&1 \\ \hline
                \end{tabular} $\cdots$
\end{center}
This implies that the length $l$ of every cluster of zeros decreases by
two every time step, i.e.
\begin{equation}
M(l,t+1)=M(l+2,t),
\end{equation}
where $M(l,t)$ denotes a number of clusters of zeros of size $s$ per
site at time $t$, and therefore
\begin{equation}
M(l,t)=M(l+2t,0).
\end{equation}
For a random initial configuration with initial density $\rho_0$, the
cluster density is given by $M(l,0)=(1-\rho_0)^l \rho_0^2$, hence
\begin{equation}
 M(l,t)=(1-\rho_0)^{l+2t} \rho_0^2.
\end{equation}
The density of zeros at time $t$,  denoted by $\eta(t)$,
is given by
\begin{equation}
\eta(t)=\sum_{l=1}^{\infty}l M(l,t) = (1-\rho_0)^{2t+1},
\end{equation}
and finally the density of ones $\rho(t)$ is equal to
\begin{equation}
\rho(t)=1-\eta(t)=1-(1-\rho_0)^{2t+1}.
\end{equation}

(ii) For Rule $238$, the derivation is similar, except that now every
cluster of zeros decreases by one every time step, which means that we
have to replace $2t$ by $t$ in the previous result, i.e. 
\begin{equation}
\rho(t)=1-(1-\rho_0)^{t+1}.
\end{equation}

(iii) In Rule 222, individuals ``dislike overcrowding'', and they do not
become adopters if their two neighbors are already adopters.
adopt if both neighbors adopted, i.e. $\set{1,0,1} \To 0$. As a
consequence, clusters of size $l>1$ decrease their size by two units
every time step, while clusters consisting of a single isolated zero
($l=1$) do not change their size. Clusters of size $1$ are thus
created from clusters of size $3$, as well as those of size $1$. For the
density of ones, this
yields (using a similar reasoning as before) 
\begin{equation}
  \rho(t)=1-\rho_0^2(1-\rho_0) - \rho_0 \frac{(1-\rho_0)^3}{2-\rho_0}
  -\left[1-\rho_0^2 + \frac{\rho_0(1-\rho_0)^2}{\rho_0-2}\right]
  (1-\rho_0)^{2t+1},
\end{equation}
and
\begin{equation}
\lim_{t \to \infty} \rho(t)= \rho(\infty)=1-\rho_0^2(1-\rho_0) - \rho_0
\frac{(1-\rho_0)^3}{2-\rho_0}.
\end{equation}

(iv) Rule 206 is just like Rule 222, except that clusters of zeros
decrease their length by one. Therefore, we obtain
\begin{equation}
 \rho(t)=1-\rho_0(1-\rho_0)-(1-\rho_0)^{t+2},
\end{equation}
and $\rho(\infty)=1-\rho_0(1-\rho_0)$.

(v) Rule 236 is the simplest because, after the
first iteration, all clusters of zeros of size $1$ disappear while
all other clusters remain unchanged, therefore
\begin{equation}
 \rho(t)=1-(1-\rho_0)^2(1+\rho_0).
\end{equation}

In summary, in all cases (except for  Rule 236), the density of ones
approaches the fixed point $\rho(\infty)$ exponentially. In a real social
system, however, this is not the case. The density of adopters usually
follows an S-shaped or logistic curve. The model discussed in the next
section eliminates this shortcoming.

\section{Probabilistic model}
In order to generalize the  simple model discussed previously, 
consider a 2-state probabilistic cellular automaton, with a dynamics
such that $s(i,t+1)$ depends on $s(i,t)$ and $\sigma(i,t)$, where
   \begin{equation}
\sigma(i,t)=\sum_{n=-\infty}^{\infty}s(i+n,t)p(n),
\end{equation}
and $p$ is a nonnegative function
 satisfying
\begin{equation}
\sum_{n=-\infty}^{\infty}p(n)=1.
\end{equation}
\citep[deterministic automata networks  of this type have been studied
in details by][]{bfg97}.

Our generalized model is defined as follows. At every
time step, a neutral individual located at site $i$ at time $t$  can
become an adopter at time $t+1$ with a
probability depending on $\sigma(i,t)$ (in this section we
will simply assume that this probability is equal to $\sigma(i,t)$). As
before, once an individual becomes an adopter, he remains an adopter,
i.e. his state cannot change.

\begin{figure}
 \myfig{0.8}{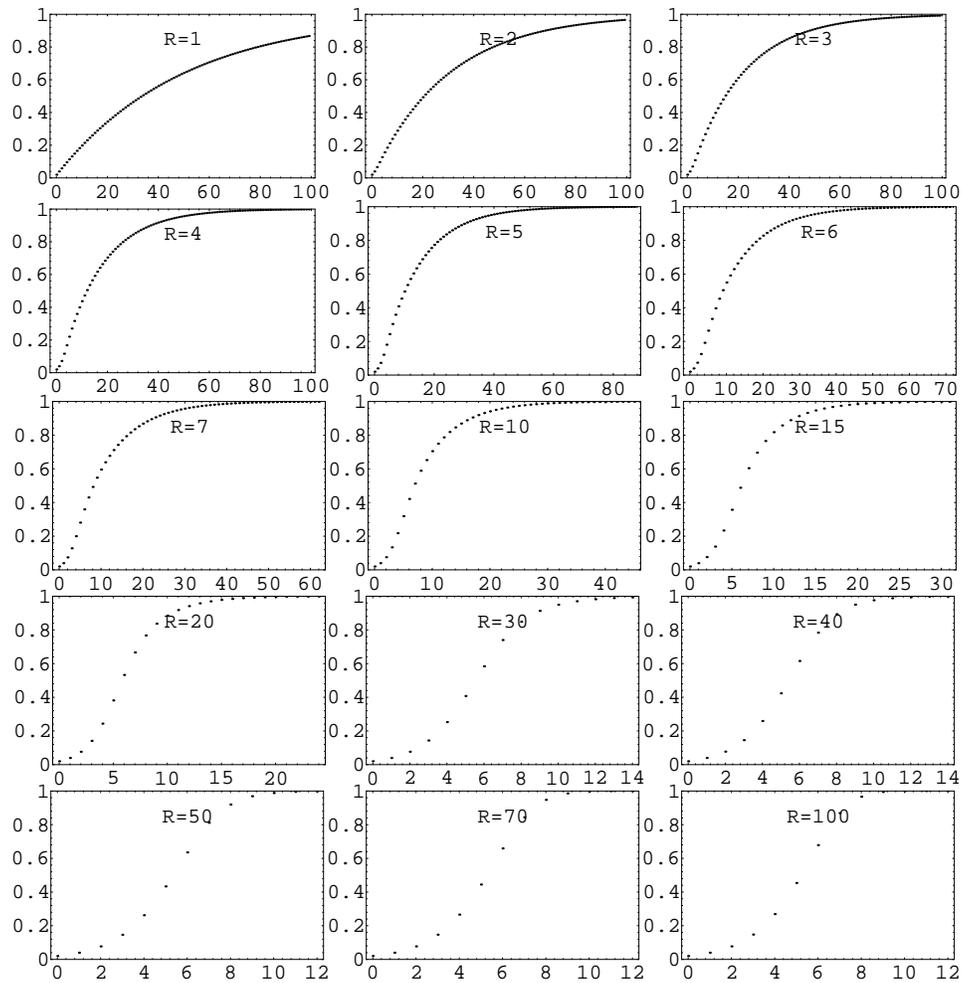}
 \caption{Density of adopters (vertical axis) as a function of time
 (horizontal axis) for several values of radius $R$.}
\label{fignoft}
\end{figure}

The model can be viewed as a probabilistic cellular automaton with the
probability distribution
\begin{eqnarray}
   P\Big(s(i,t+1)=0\Big)&=&\Big(1-s(i,t)\Big)\Big(1-\sigma(i,t)\Big) \\
   P\Big(s(i,t+1)=1\Big)&=&1-\Big(1-s(i,t)\Big)\Big(1-\sigma(i,t)\Big)
\end{eqnarray}
The {\it transition probability} $P_{b \leftarrow a}$ is defined as
\begin{equation}
P_{b \leftarrow a}=P\Big(s(i,t+1)=b | s(i,t)=a\Big),
\end{equation}
and represents the probability for a given site of changing its state
from $a$ to $b$ in one time step. In our case, the transition
probability matrix has the form
\begin{equation}
 \mathbf{P}=
 \left[
 \begin{array}{cc}
  P_{0 \leftarrow 0} & P_{0 \leftarrow 1} \\
  P_{1 \leftarrow 0} & P_{1 \leftarrow 1}
 \end{array}  \right] =
 \left[
 \begin{array}{cc}
  1-\sigma(i,t) & 0 \\
    \sigma(i,t) & 1
 \end{array} \right] .
 \label{probmatrix}
\end{equation}
As a first approximation we we consider $\sigma(i,t)$ defined by
\begin{equation}
 \sigma(i,t)=\frac{1}{2R}\left(\sum_{n=-R}^{-1}s(i+n,t)+
 \sum_{n=1}^{R}s(i+n,t)\right).
\end{equation}
$\sigma(i,t)$ is then the local density of adopters at time
$t$ over the $2R$ closest
neighbors of site $i$. This choice of $\sigma$, although somewhat
simplistic,
captures some essential features of a real social system: the number
of influential neighbors is finite and these neighbors are all located
within a certain finite radius $R$. Opinions of all neighbors have
equal weight here, which is maybe not realistic, but good enough as a
first approximation. Let $\rho(t)$ be the global density of adopters at
time $t$ (i.e. number of adopters per lattice site), and
$\eta(t)=1-\rho(t)$. Since  $P_{1 \leftarrow 1}=1$, the number of
adopters increases with time, and $\lim_{t \rightarrow \infty}
\rho(t)=1$. If we start with a small initial density of randomly
distributed adopters $\rho_0$, $\rho(t)$ follows a characteristic
S-shaped curve, typical of many growth processes. The curve becomes
steeper when $R$ increases, and if $R$ is large enough it takes only a
few time steps to reach a high density of ones (e.g. $0.99$). Figure
\ref{fignoft} shows some examples of curves obtained in a computer
experiment with a lattice size equal to $10^5$ and $\rho_0=0.02$.

\section{Exact Results}

Average densities $\rho(t+1)$ and $\eta(t+1)$ can be obtained from
previous densities $\rho(t)$ and $\eta(t)$ using the transition probability
matrix ( $\langle \rangle$ denotes a spatial average)
\begin{equation}
\left[
\begin{array}{c}
 \eta(t+1)  \\
 \rho(t+1)
\end{array}
\right]
=
\left[
\begin{array}{cc}
 \langle P_{0 \leftarrow 0} \rangle & \langle P_{0 \leftarrow 1} \rangle \\
 \langle P_{1 \leftarrow 0} \rangle & \langle P_{1 \leftarrow 1} \rangle
\end{array}  \right]
\left[
\begin{array}{c}
 \eta(t)  \\
 \rho(t)
\end{array}
\right],
\end{equation}
hence
\begin{equation}
\rho(t+1)=\rho(t) \langle P_{1 \leftarrow 1} \rangle + (1-\rho(t))
\langle P_{1 \leftarrow 0} \rangle,
\end{equation}
and using (\ref{probmatrix}) we have
\begin{equation}
\rho(t+1)=\rho(t) + (1-\rho(t)) \langle \sigma(i,t) \rangle. \label{difeq}
\end{equation}
This difference equation can be solved in two special cases,
$R=1$ and $R=\infty$.

If $R=1$, only three possible values of local density $\sigma$ are
allowed: $0,\frac{1}{2}$ and $1$. The initial configuration can be
viewed as clusters of ones separated by clusters of zeros. Only
neutral sites adjacent to a cluster of ones can change their state,
and they will become adopters with probability $\frac{1}{2}$. All
other neutral sites have a local density equal to zero, therefore they
will remain neutral. This implies that the length of a cluster of
zeros will on the average decrease by one every time step, just like in
the case of Rule 238. The density of adopters, therefore, is, as for Rule
238, given by
\begin{equation}
\rho(t)=1-(1-\rho_0)^{t+1}.
\end{equation}
Using the above expression, we obtain
\begin{equation}
\rho(t+1)=\rho(t) + (1-\rho(t)) \rho_0.
\end{equation}
Comparing with (\ref{difeq}) this yields $\langle \sigma(i,t) \rangle
=
\rho_0$, i.e. the average probability that a neutral individual
adopts the innovation is time-independent and equal to the initial
density of adopters, which was {\it a priori\/} not obvious.

When $R \rightarrow \infty$, the local density of ones becomes equivalent
to the global density, thus in (\ref{difeq}) we can replace $\langle
\sigma(i,t) \rangle$ by $\rho(t)$. Hence
\begin{equation}
\rho(t+1)=\rho(t) + (1-\rho(t))\rho(t),
\end{equation}
or
\begin{equation}
1-\rho(t+1)=(1-\rho(t))^2,
\end{equation}
that is
\begin{equation}
\rho(t)=1-(1-\rho_0)^{2^t}.
\end{equation}
Note that this case corresponds to the mean-field approximation, as we
neglect all spatial correlations and replace the local density by the
global one.

For $1<R <\infty$, we may assume that the density of adopters can be
written in the form
\begin{equation}
\rho(t)=1-(1-\rho_0)^{f(t,R)}, \label{fdef}
\end{equation}
where $f$ is a certain function satisfying $f(t,1)=t+1$ and $f(t,
\infty )=2^t$. Computer simulations suggest that for a finite
$R$, $f(t,R)$ becomes asymptotically linear (when $t \rightarrow
\infty$), and the slope of the asymptote increases with $R$.
Moreover, for large $R$, $f$ satisfies the following approximate
equation
\begin{equation}
f(t+k, 2^k R)=2^k f(t,R), \label{scaling}
\end{equation}
where k is a positive integer. Detailed discussion of $f(t,R)$ and its
properties can be found in \citet{fb96}.

\section{Generalization}
The model presented in the previous chapter was rather crude. One of
its assumptions is that once the individual accepts the innovation,
he will never change his mind. In practice, every technology or product
has a finite life span. For some products, as TV sets, this time is
relatively long, while for other items, like computer software, it is
much shorter. One way to incorporate this phenomenon in our model is
to assume that at every time step, any adopter can drop the innovation
with a given probability $p$. 
Therefore, the average time during which an individual is an adopter is
$1/p$ (geometric distribution). 

To make the model even more realistic, we will also assume that the
adoption probability is not equal but proportional to the local density of
adopters
\begin{equation}
 P_{1 \leftarrow 0}=q \sigma(i,t),
\end{equation}
where $q \in [0,1]$. Hence, the new transition
probability matrix is
\begin{equation}
 P_{mn}=
\left[
\begin{array}{cc}
 P_{0 \leftarrow 0} & P_{0 \leftarrow 1} \\
 P_{1 \leftarrow 0} & P_{1 \leftarrow 1}
\end{array}  \right] =
\left[
\begin{array}{cc}
 1-q \sigma(i,t) & p \\
   q \sigma(i,t) & 1-p
\end{array} \right] .
\end{equation}
The difference equation for the average density of adopters
\begin{equation}
\rho(t+1)=\rho(t) \langle P_{1 \leftarrow 1} \rangle + \Big(1-\rho(t)\Big)
\langle P_{1 \leftarrow 0} \rangle
\end{equation}
now becomes
\begin{equation}
\rho(t+1)=(1-p)\rho(t) + q\Big(1-\rho(t)\Big) \langle \sigma(i,t) \rangle.
\label{difeqdecay}
\end{equation}
The previous model is recovered for $p=0$ and
$q=1$. When $q=0$, the solution is $\rho(t)=\rho_0(1-p)^t$. When $R=1$,
the model can also be considered as a discrete version of the contact
process, an irreversible lattice model involving nearest-neighbor
interactions, used to study catalytic reactions
\citep{harris74,liggett85}. In the contact process, a particle desorbs
spontaneously with rate $p$ and adsorbs at a given unoccupied site at
a rate proportional to the number of neighboring occupied sites. 

Although it is not possible to solve exactly equation (\ref{difeqdecay}),
the general nature of the solution can be understood using
the mean-field approximation (MFA), in which it is assumed
that the average local density $\langle
\sigma(i,t) \rangle$ is the same as the global density $\rho(t)$ (this
is actually true for $R=\infty$). Our equation becomes
\begin{equation}
\rho(t+1)=(1-p)\rho(t) + q \rho(t) \Big(1-\rho(t)\Big).
\end{equation}
The substitution
\begin{equation}
\rho(t)=\frac{1-p+q}{q}x(t)
\end{equation}
yields
\begin{equation}
x_{t+1}=(1-p+q)x(t) \Big(1-x(t)\Big).
\end{equation}
For $R=\infty$, therefore, the dynamics of the model can be understood
as an iteration of the logistic map $Q_\lambda: x \mapsto \lambda x
(1-x)$ with $\lambda=1-p+q$. Note that $0 \leq \lambda \leq 2$, which
excludes stable periodic points and chaos.

$Q_\lambda$ has always two fixed points $x^{(1)}(\infty)=0$ and
$x^{(2)}(\infty)=1-1/\lambda$. Only one, however, is stable, depending
on $\lambda$, namely  $x^{(1)}(\infty)$ when $\lambda < 1$ and
$x^{(2)}(\infty)$ when $\lambda>1$. In terms of $\rho(t)$ we obtain
\begin{equation}
 \lim_{t \rightarrow \infty} \rho(t)= \left \{ \begin{array}{ll}
                       0 & \mbox{if $p>q$,}\\
                     1-p/q & \mbox{otherwise.}
                    \end{array}
 \right.
\end{equation}
Since  $|Q_\lambda^\prime (x_\infty)|<1$ if $\lambda \neq 1$, the stable
fixed
point is hyperbolic (strongly attracting). At $\lambda =1$, however,
it becomes nonhyperbolic (weakly attracting), and $Q_\lambda$ exhibits a
\ndef{transcritical bifurcation} \citep{Bardos80} with exchange of
stability.

\section{Phase Transition}

\begin{figure}[t!]
\myfig{1.0}{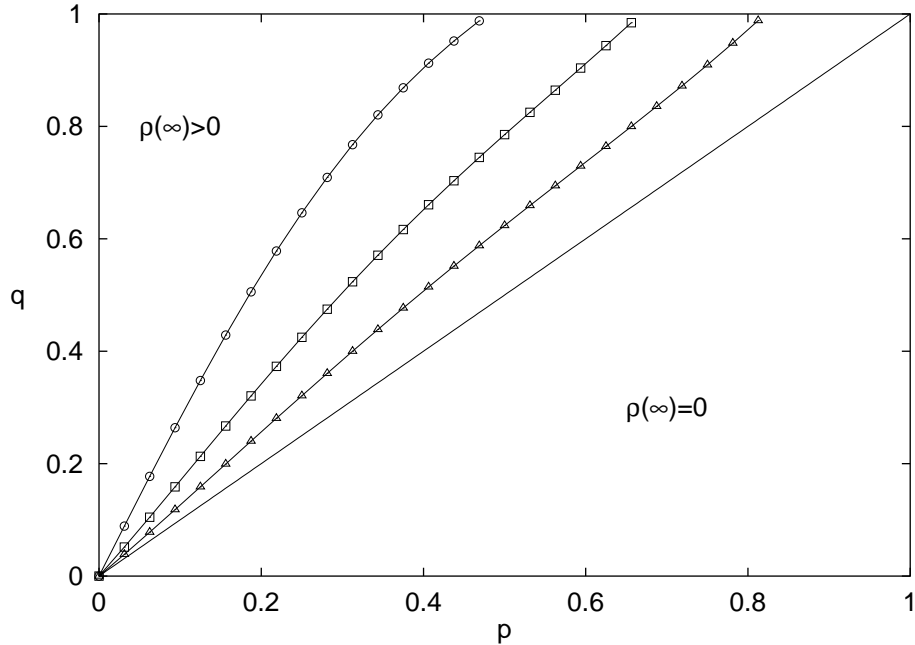}
\caption{$(p,q)$ phase diagram obtained using Monte Carlo simulations.
Lines separating regions  where $\rho(\infty)=0$ and $\rho(\infty)
\neq 0$ are shown for $R=1$ ($\circ$), $R=4$ ($\Box$), and $R=16$
($\triangle$). The dotted line represents the mean-field
approximation. Lattice size is equal to $10000$.}
\label{phased}
\end{figure}

\begin{figure}[t!]
\myfig{1.0}{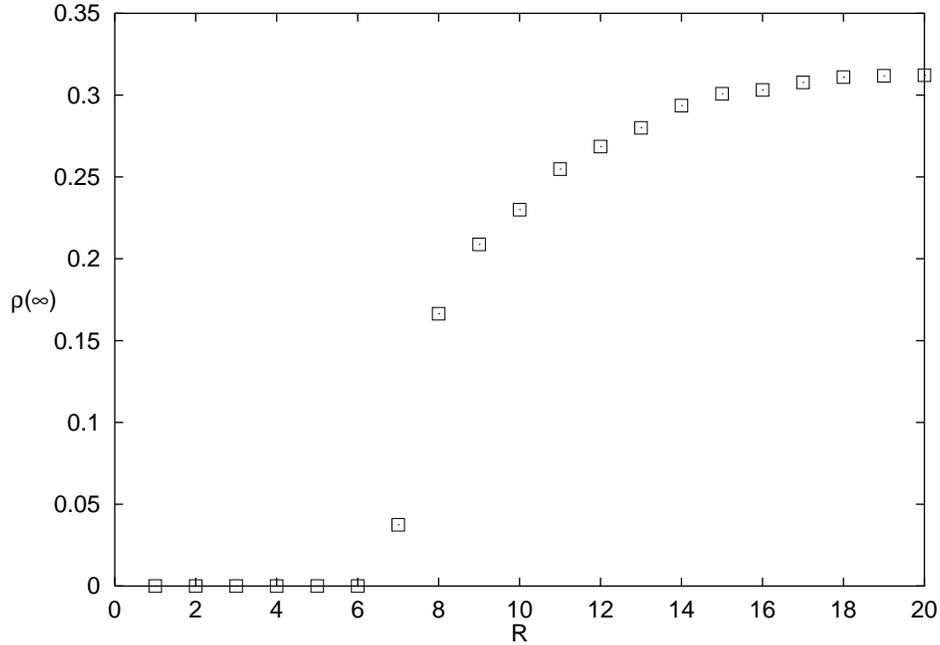}
\caption{Phase transition of $\rho(\infty)$ with $R$ playing the role of the
control parameter for $p=0.2$ and $q=0.3$.}
\label{rbifur}
\end{figure}

The mean-field approximation discussed in the previous section becomes
correct only when $R \rightarrow \infty$. For small values of $R$,
like in the basic model, strong correlations are created and
substantial deviations from MFA can be expected. Figure \ref{phased}
represents the $(p,q)$ phase diagram for different values of $R$
obtained in computer simulations. The smaller $R$ is, the larger the
deviation from MFA (dotted line) becomes.
As we can see, the line separating $\rho(\infty) \neq 0$ and
$\rho(\infty)=0$ shifts to the left (toward larger $p$) when $R$
increases. This means that the connectivity between sites increases
the robustness of the innovation, i.e. even if $p$ is large, it can be
compensated by a large $R$. To see it, consider a point on the
$(p,q)$ phase diagram which is located between the $R=1$ and the
mean-field
lines (e.g. $(0.2, 0.3)$). If we plot $\rho(\infty)$
as a function of $R$, we
obtain a bifurcation (or phase-transition) diagram, as shown
in Figure \ref{rbifur}. For $R\leq 6$ the asymptotic density of
adopters goest to zero, but if
$R>6$, $\rho(\infty)$ becomes positive. Three distinct
regions of the $(p,q)$ phase space, therefore, can be distinguished:
\begin{enumerate}
\item In the region bounded by the $q$ axis and the $R=1$ critical line,
 $\rho(\infty)>0$ regardless of $R$.
\item In the region bounded by the $R=1$ and the mean-field critical
lines,
$\rho(\infty)$ is either equal to zero or positive, depending on the value
of $R$. In general, $\rho(\infty)$ is positive for a 
sufficiently large $R$.
\item In the region bounded by the mean-field critical line,
$\rho(\infty)=0$ regardless of $R$.
\end{enumerate}

\section{$R=1$ critical line}

\begin{figure}[t!]
 \myfig{1.0}{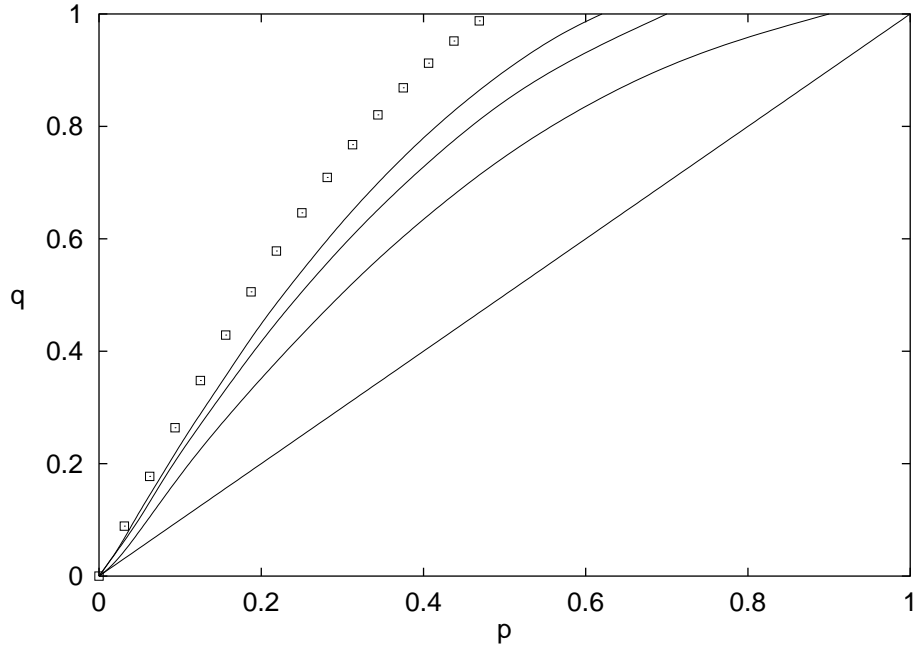}
\caption{$R=1$ critical line ($\Box$) and its local structure theory
approximations. Starting from the right, the consecutive solid lines
correspond to $n=1$, $2$, $3$ and $4$.}
\label{mfa}
\end{figure}

We studied the boundary between regions $1$ and $2$, i.e., the $R=1$
critical line, using the local structure theory (LST) up to order
$4$ \citep{gutowitz87,gutowitz87a}, as shown in Figure \ref{mfa}.

\begin{figure}[t!]
 \myfig{0.58}{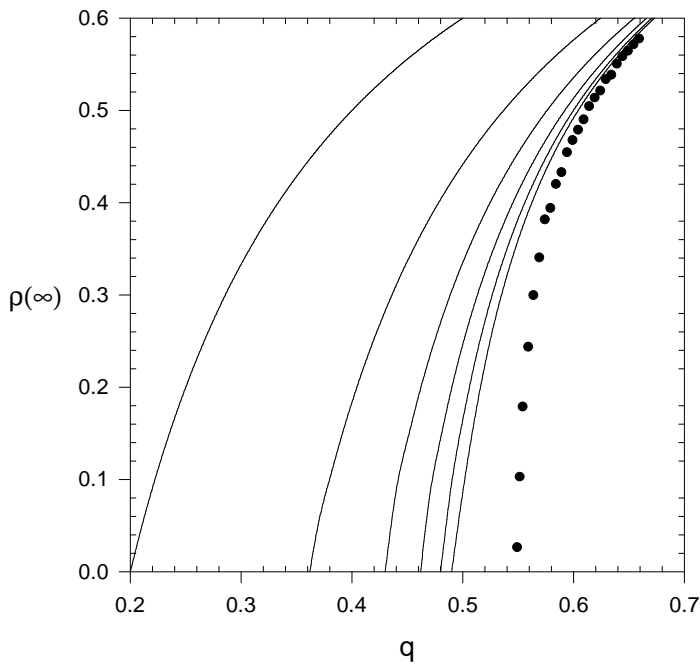}
\caption{Local structure approximation for $p=0.2$. Solid lines
represent consecutive orders of the local structure approximations,
starting with $n=1$ (the leftmost curve), and ending with $n=6$ (the
rightmost solid curve). The dots represent results of computer
simulations ($p=0.2$, $R=1$).}
\label{lstg}
\end{figure}

As we increase the order $n$ of the approximation, we obtain an
approximate
critical line closer to the experimental one.
This can be also seen if we
plot $\rho(\infty)$ obtained from LST as a function $p$, keeping $q$
constant. An example of such curves is shown in Figure \ref{lstg},
where six consecutive orders of LST are presented. For this particular
value of $p$, the sixth-order critical $q$-value $q_c$ differs by about
$11\%$ from the numerical result, as shown below.
\begin{center}
\begin{tabular}{|c|c|c|c|c|c|c|c|} \hline
$n$   & 1     &  2    &   3   &   4    &  5     &  6    & simulation \\ \hline
$q_c$ & 0.200 & 0.362 & 0.430 & 0.462  & 0.479  & 0.490 & 0.549 \\ \hline
\end{tabular}
\end{center}

As mentioned earlier, for $q=1$ and $R=1$, our model can be
considered as a discrete realization of the contact process. For the
contact process, theoretical studies as well as Monte Carlo
simulations suggest that the phase transition occurs at $p_c$=0.3032
\citep{dickman91,jensen93}. The table below compares this result with
our numerical simulations and local structure theory approximate results.
\begin{center}
\begin{tabular}{|c|c|c|c|c|c|c|c|} \hline
$n$   & 1     &  2    &   3   &   4    &  5    & simulation &CP \\ \hline
$p_c$ & 1.00 & 1.00 & 0.70 & 0.62  & 0.55 & 0.48 &0.3032\\ \hline
\end{tabular}
\end{center}
The discrepancy between discrete and continuous time versions of the
contact process is understandable, since the synchronous dynamics of
CA allows, for example, for one site to change its state from 1 to 0,
and, at {\it exactly} the same time, for its neighbor to change  its
state from 0 to 1, something that does not occur in the continuous time
model. We should stress at this point that although there are many
analytical techniques developed to treat stochastic processes such as
the contact process, none of them, in general, can be easily
translated to the language of cellular automata theory, and the
synchronicity of updating is usually the main source of difficulties.
For example, the cluster approximation recently proposed for the
contact process \citep{BenNaim96} belong to this category.

On the other hand, when all transition probabilities are small, a
single site of the lattice is rarely changed, and during most of the
iterations of the CA it will remain in the same state. The updating,
therefore, becomes more and more similar to  asynchronous updating,
and as a consequence, we can expect that the model should behave
almost like the time-continuous contact process.

\section{Conclusion}
We have studied a probabilistic cellular automata model for the spread
of innovations. Our results emphasize the importance of the range of the 
interaction between individuals. The innovation spreads faster
when the range increases since increased connectivity between individuals
reduces constrains on the exchange of information. 
Larger connectivity could be also achieved by increasing space
dimensionality.

The range of interaction $R$ not only affects the growth rate, but in
the region of parameter space $(p,q)$  bounded by the $R=1$ critical line
and the mean-field line it can be a decisive factor for the asymptotic
density of adopters $\rho(\infty)$. In this region, if $R$ is too small,
$\rho(\infty)=0$, while for a sufficiently large connectivity
$\rho(\infty)>0$.

\end{document}